# Excitation Energy Dependent Raman Signatures of ABA- and ABC-stacked Few-layer Graphene

The An Nguyen[1], Jae-Ung Lee[1], Duhee Yoon[1]†, and Hyeonsik Cheong[1]*

[1]Department of Physics, Sogang University, Seoul 121-742, Korea.

†Current address: Cambridge Graphene Centre, Engineering Department, University of Cambridge, Cambridge, CB3 0FA, United Kingdoms.

**The dependence of the Raman spectrum on the excitation energy has been investigated for ABA-and ABC- stacked few-layer graphene in order to establish the fingerprint of the stacking order and the number of layers, which affect the transport and optical properties of few-layer graphene. Five different excitation sources with energies of 1.96, 2.33, 2.41, 2.54 and 2.81 eV were used. The position and the line shape of the Raman 2D, G$^*$, N, M, and other combination modes show dependence on the excitation energy as well as the stacking order and the thickness. One can unambiguously determine the stacking order and the thickness by comparing the 2D band spectra measured with 2 different excitation energies or by carefully comparing weaker combination Raman modes such as N, M, or LOLA modes. The criteria for unambiguous determination of the stacking order and the number of layers up to 5 layers are established.**

Few-layer graphene (FLG) is attracting much interest recently. The electronic band structure is sensitive to interlayer interactions which in turn depend on the number of layers and stacking orders. In natural graphite, two stable stacking orders, ABA or Bernal stacking and ABC or rhombohedral stacking, exist[1–3]. Even for the same number of layers, different stacking sequences result in different physical properties because of the interlayer interactions. Several studies have elucidated the influence of the number of layers and the stacking order on transport[4,5] and optical properties[6–10]. These results suggest that the number of layers and stacking orders are crucial to understand the properties of the FLG system. Therefore, identifying these structural differences is very important. Supplementary Fig. S1 summarizes the phonon modes in ABA- and ABC-stacked 3-layer graphene (3LG). Similar assignments can be made for thicker FLG. The stacking order affects the phonon dispersion through different interlayer interaction. Raman spectroscopy has been established as the most powerful tool to determine the structure of FLG nondestructively. The line shape of the Raman 2D band has been used to determine the number of layers[6–8,10,11] and stacking orders[9,12]. Early studies[6–8,10,11] established Raman fingerprints to distinguish the number of layers in FLG using the 514.5 or 532-nm lasers without specific reference to the stacking order. A later study by Lui *et al.* showed that Raman imaging technique can be used to visualize the ABA and ABC domains on 3LG and 4LG samples.[9] Because of the double-resonance Raman scattering process for the 2D band, its line shape reflects not only the phonon dispersion but also the electronic band structure, which in turn are determined by the stacking order and the number of layers. However, as we show in this work, the line shapes of the 2D band from FLG samples with different stacking orders or numbers of layers are sometimes very similar for a given excitation energy, making it difficult to determine these structural parameters unambiguously. On the other hand, the low frequency shear mode of FLG, the so-called C peak,[13] directly reflects the interlayer coupling and so shows a clear correlation with the number of layers. However, because of its proximity to the laser line, it is very difficult to observe with conventional confocal Raman apparatuses. Recently, several studies demonstrated that other resonant Raman modes, such as so-called N mode[14,15], M mode[14–18], or other combinational modes[17,18] also reflect the stacking order and the number of layers. One may, therefore, determine the stacking order and the number of layers unambiguously, by combining information from several such modes. It should be pointed out, however, that these resonant modes are dispersive because the momenta of the phonons involved are dependent on the excitation energy. Therefore, it is essential to have comprehensive understanding of the dependences of these Raman features on the excitation energy as well as the stacking order and the number of layers. As different researchers use different lasers for the Raman measurements, the excitation energy dependence is a crucial piece of information for using the Raman spectrum as the fingerprint of the stacking order and the number of layers.

In this work, we used five most popular excitation laser wavelengths (632.8, 532, 514.5, 488, and 441.6 nm, corresponding to 1.96, 2.33, 2.41, 2.54, and 2.81 eV, respectively) to investigate various Raman features of FLG. It is found that typically 2 different excitation energies are needed to determine the structure by using only the 2D band. The G$^*$ band is quite useful for differentiating stacking orders if a high-energy excitation laser is used. Other weaker modes also depend sensitively on the structural parameters if an appropriate excitation energy is used. Our work shows how the stacking order and the number of layers can be determined unambiguously from Raman measurements.

**Results**

**2D band.** The 2D band, also known as the G' band, originates from the double-resonant process which involves two transverse optical (TO) phonons near the K points in the Brillouin zone[19] Figure 1 shows the Raman 2D bands of FLG with ABA- and ABC- stacking orders under 5 different excitation energies. The excitation energy dependence of the 2D mode without regard to the stacking order has been studied by several groups[6,7,20]. Cong *et al.*[12] and Lui *et al.*[9] studied ABA- and ABC-stacked 3LG with different excitation energies. In our work, we investigated the excitation energy dependence for 1 to 5 layer graphene (1LG - 5LG) with both ABA- and ABC-stacking. The blueshift of the 2D bands with increasing excitation energy is explained by the resonant scattering process which involves phonons with momenta in the ΓK direction[21,22]. The line shape of the 2D band reflects the contributions of scattering paths involving different valence and conduction bands, and so the frequency of the components reflect the details of the band structure near the given excitation energy. Furthermore, the electron-phonon interaction could in principle depend on the energy and momenta of the electrons and phonons, which would lead to excitation energy dependence of the relative intensities of different components of the 2D band.

Due to the dramatic influence of the number of layers and the stacking order on the electronic band structure, the line shape of the 2D band exhibits clear dependence on the number of layer, the stacking order, and the excitation energy and therefore can be used as the fingerprint of the number of layers and the stacking order if appropriate excitation energy is chosen. For example, with the 2.54 eV excitation, most cases can be determined with a possible exception of ABA-stacked 4 (4LG) and 5 layer graphene (5LG). In order to distinguish them, one would need to use a lower energy excitation source such as 1.96 or 2.33 eV. On the other hand, if one uses the 1.96 eV excitation, all but the ABC-stacked 3 and 4 layer graphene can be identified. To distinguish them, one needs to use higher energy excitations such as 2.54 or 2.81 eV.

**G\* band.** The G\* band which appears in the 2400–2450 cm$^{-1}$ region of the spectra is interpreted as a combination of a transverse optical (TO) and longitudinal acoustic (LA) phonon mode near the K point resulting from double resonance Raman scattering processes[23] and is commonly denoted as D+D". Figure 2 shows the G\* band of our graphene samples. It has an asymmetric line shape which can be fitted with 2 Lorentzian components. It has been suggested that the main peak originates from phonons along the KΓ high symmetry line and that the asymmetry is due to phonons with momenta away from that high symmetry line[23]. This band shifts to the lower frequency as the excitation energy is increased from 1.96 to 2.81 eV.

Figure 2 clearly shows the dependence of the line shape on the number of layers and the stacking order. The dependence is stronger for higher excitation energies. For example, with the 1.96 eV excitation, it is difficult to distinguish 1 to 3 layer graphene, and the difference between ABA- and ABC stacked FLG is minimal. However, with the 2.81 eV excitation, single- and bi-layer graphene can be easily identified with respect to thicker FLG. Furthermore, ABA- and ABC-stacking can be distinguished for 3 to 5 layer graphene, although for the same stacking, the number of layers cannot be determined conclusively. In general, the ABC-stacked FLG has smaller intensity on the lower frequency side of the G* band. The lower frequency component is due to phonons along the KΓ high symmetry line according to Ref. 23. Therefore, its smaller intensity can be interpreted as being due to weaker electron-phonon coupling of this particular phonon mode in ABC-stacked FLG.

**M band.** In the range of 1700 to 1800 cm$^{-1}$, a series of Raman peaks, called the M band, appear as shown in Fig. 3. The intensity and the number of peaks depend strongly on the stacking order and the number of layers. These peaks have been interpreted as combinations of an LO and an out-of-plane breathing mode ZO' (LO+ZO') and an overtone of another out-of-plane breathing mode ZO (2ZO)[14,18]; or as overtones of out-of-plane transverse optical oTO modes (2oTO's)[16,17]. This band is not observed for 1LG as other researchers reported[14,16–18]. For the ABA-stacked FLG, a sharp peak dominates with smaller side peaks. The line shape does not change substantially with the number of layers for 3 or more layers. For the ABC-stacked FLG, on the other hand, there are several sharp peaks, and the number of peaks increases with the number of layers. So this band can be used to distinguish ABA and ABC stackings and also to identify the number of layers for ABC-stacked FLG. The positions of these sharp peaks have definitive dependence on the excitation energy as shown in Fig. 4. At lower energy excitations, the peak separations are larger, and so it is easier to distinguish the stacking orders.

**Weak combination modes (1780 – 2250 cm$^{-1}$).** Recently, several weak features have been discovered in the Raman spectra of FLG in the range between 1780 and 2250 cm$^{-1}$ [14,16–18]. These features originate from combinations of Raman modes which result from double-resonance Raman scattering processes involving LO, LA, in-plane transverse optical (iTO), and in-plane transverse acoustic (iTA) modes. These peaks are shown in Fig. 5.

For the same type of stacking the spectrum does not vary much with the number of layers. On the other hand, there is a clear difference between the spectra for ABA-stacked FLG and ABC-stacked FLG. The LOLA mode in particular can be used as a fingerprint of the stacking. This mode is sharper and appears at higher frequencies in ABC-stacked FLG. This is similar to what

Cong et al. observed for 3LG[12]. The iTOTA (iTO+iTA) mode also shows different line shapes for the two stackings, although the difference is rather subtle. These trends are observed for all excitation energies used in our study (See Supplementary Fig. S7 online).

The correlation between the peak position of the combination modes from 1780 to 2250 cm$^{-1}$ and the number of layers in ABA- and ABC-stacked graphene under different excitation energies is summarized in Fig. 6. As the excitation energy increases from 1.96 to 2.81 eV, the iTALO (1854 – 1867 cm$^{-1}$) and LOLA peaks (2011 – 2018 cm$^{-1}$) blueshifts. On the contrary, the iTOTA mode redshifts with increasing excitation energy. From the excitation energy dependence we can get insight upon the scattering mechanism. For intra-valley scattering, the momenta of phonons are close to the Γ point and increase as the excitation energy is increased. When an acoustic phonon is involved, the strong dispersion of the acoustic phonons near the Γ point would lead to a blueshift of the Raman features with increasing excitation energy. This is what is seen for LOLA and iTALO modes. For inter-valley scattering, on the other hand, higher excitation energy would lead to smaller phonon momenta, moving the momenta of the phonons away from the K point. For the iTOTA mode, the acoustic phonon frequency decreases, whereas the iTO frequency increases because of the Kohn anomaly, as the momenta moves away from the K point. Since iTA has a steeper dispersion near the K point than iTO, the overall effect should be a redshift, as seen in Fig. 6.

**N band.** On the lower frequency side of the G peak (~1580 cm$^{-1}$), a weak feature, called the N band is observed. Herziger et al. attributed these peaks to a combination mode of a Stokes-scattered LO phonon and an anti-Stokes scattered ZO' phonon[14]. The stacking-order dependence of these peaks was not investigated. Figure 7 shows the N band for ABA- and ABC-stacked FLG for 5 different excitation energies. The band redshifts with increasing number of layers. As the excitation energy increases the difference between ABA- and ABC-stacking becomes more dramatic, whereas the peaks are better resolved with lower energy excitations. Low-energy excitations can be used to determine the number of layers regardless of the stacking order, whereas high-energy excitations should be used to distinguish the stacking order.

**Discussion**

Although the 2D band is very sensitive to both the number of layers and the stacking order, a single excitation energy is not sufficient to provide a definitive fingerprinting for ABA- and ABC stacked FLG up to 5 layers in thickness. By combining measurements of the 2D band spectra with 2 different excitation sources, a low energy laser and another high energy one, we can definitively determine the stacking order and the number of layers of FLG. Additionally, a distinction between ABA- and ABC-stacking is revealed in the G* band for high-energy excitations. Weak combination modes such as N, M, and LOLA bands also exhibit clear dependence on the structure of FLG and can be used to confirm the identifications of the stacking order and the number of layers. For example, the LOLA-mode peak, which blueshifts with excitation energy in the case of ABC-stacked FLG only, is useful for determination of the number of layers for ABC-stacked FLG. In conclusion, we have shown that Raman spectroscopy, with appropriate choice of excitation energy, can be used to determine unambiguously the stacking order and the number of layers of FLG up to 5 layers.

We should note that doping of the graphene samples may affect their Raman features. It is well known that the G band has a strong dependence on the doping density[24,25]. Therefore, the weak Raman features that we use to determine the number of layers and the stacking order may be affected by the residual doping of the graphene samples. However, we did not observe any appreciable difference in these Raman features among several samples that we measured. We hence conclude that the uncertainty in the determination of the number of layers and the stacking order due to residual doping is minimal.

**Methods**

Few-layer graphene samples were prepared on Si substrates covered by a layer of 100 nm oxide ($SiO_2$) by the mechanical exfoliation method. The number of layers was identified by optical microscope images, atomic force microscopy, and by the line shape of the Raman 2D band. Stacking orders were identified by comparing the Raman spectra with previously reported data in the literature[9,12,18]. Areas with different stacking were identified by the Raman imaging technique (See Supplementary Information).

For the Raman measurements, we used five different excitation sources: the 441.6-nm (2.81 eV) line of a He-Cd laser, the 488-nm (2.54 eV) and 514.5-nm (2.41 eV) lines of an Ar ion laser, the 532-nm (2.33 eV) line of a diode-pumped frequency-doubled solid state Nd:YAG laser, and the 632.8-nm (1.96 eV) line of a He-Ne laser. The laser beam was focused onto the sample by a 50× microscope objective lens (0.8 N.A.), and the scattered light was collected and collimated by the same objective. For the 441.6-, 488-, 514.5- and 532-nm excitation wavelengths, the scattered signal was dispersed with a Jobin-Yvon Triax 550 spectrometer (1800 grooves/mm) and detected with a liquid-nitrogen-cooled back-illuminated charge-coupled-device (CCD) detector. For the 632.8-nm excitation laser wavelength, the scattered signal was dispersed with a Jobin-Yvon Triax 320 spectrometer (1200 grooves/mm) and detected with a thermoelectrically cooled, back-illuminated, deep-depletion CCD detector. For low-intensity peaks such as the N band, long integration times up to 40 minutes were used in order to get spectra with good signal-to-noise ratios. The laser power was kept below 1 mW to avoid local heating and damage to the samples. For atomic force microscopy, we use a commercial system (NT-MDT NTEGRA Spectra).

**References**

1. Latil, S. & Henrard, L., Charge Carriers in Few-Layer Graphene Films. *Phys. Rev. Lett*. 97, 036803 (2006).

2. Guinea, F., Castro Neto, A. H. & Peres, N. M. R., Electronic states and Landau levels in graphene stacks. *Phys. Rev. B* 73, 245426 (2006).

3. Aoki, M. & Amawashi, H., Dependence of band structures on stacking and field in layered graphene. *Solid State Commun*. 142, 123–127 (2007).

4. Bao, W. *et al.*, Stacking-dependent band gap and quantum transport in trilayer graphene. *Nature Phys*. 7, 948–952 (2011).

5. Zou, K., Zhang, F., Clapp, C., MacDonald, A. H. & Zhu, J., Transport Studies of Dual-Gated ABC and ABA Trilayer Graphene: Band Gap Opening and Band Structure Tuning in Very Large Perpendicular Electric Fields. *Nano Lett*. 13, 369–373 (2013).

6. Ferrari, A. C. *et al.*, Raman Spectrum of Graphene and Graphene Layers. *Phys. Rev. Lett*. 97, 187401 (2006).

7. Gupta, A., Chen, G., Joshi, P., Tadigadapa, S. & Eklund, P. C., Raman Scattering from High-Frequency Phonons in Supported n-Graphene Layer Films. *Nano Lett.* 6 (12), 2667-2673 (2006).

8. Graf, D. *et al.*, Spatially Resolved Raman Spectroscopy of Single- and Few-Layer Graphene. *Nano Lett*. 7 (2), 238-242 (2007).

9. Lui, C. H. *et al.*, Imaging Stacking Order in Few-Layer Graphene. *Nano Lett*. 11, 164-169 (2011).

10. Ni, Z. H. *et al.*, Graphene Thickness Determination Using Reflection and Contrast Spectroscopy. *Nano Lett*. 7 (9), 2758-2763 (2007).

11. Yoon, D. *et al.*, Variations in the Raman Spectrum as a Function of the Number of Graphene Layers. *Journal of the Korean Physical Society* 55 (3), 1299-1303 (2009).

12. Cong, C. *et al.*, Raman Characterization of ABA- and ABC-Stacked Trilayer Graphene. *ACS Nano* 5 (11), 8760–8768 (2011).

13. Tan, P. H. *et al.*, The shear mode of multilayer graphene. *Nat. Mater.* 11, 294-300 (2012).

14. Herziger, F., May, P. & Maultzsch, J., Layer-number determination in graphene by out-of-plane phonons. *Phys. Rev. B* 85, 235447 (2012).

15. Herziger, F. & Maultzsch, J., Influence of the layer number and stacking order on out-of-plane phonons in few-layer graphene. *Phys. Status Solidi B* 250 (12), 2697-2701 (2013).

16. Cong, C., Yu, T., Saito, R., Dresselhaus, G. F. & Dresselhaus, M. S., Second-Order Overtone and Combination Raman Modes of Graphene Layers in the Range of 1690−2150 cm$^{-1}$. *ACS Nano* 5, 1600–1605 (2011).

17. Rao, R. *et al.*, Effects of Layer Stacking on the Combination Raman Modes in Graphene. *ACS Nano* 5 (3), 1594–1599 (2011).

18. Lui, C. H. *et al.*, Observation of Layer-Breathing Mode Vibrations in Few-Layer Graphene through Combination Raman Scattering. *Nano Lett*. 12 (11), 5539−5544 (2012).

19. Malard, L. M., Pimenta, M. A., Dresselhaus, G. & Dresselhaus, M. S., Raman spectroscopy in graphene. *Phys. Rep*. 473, 51-87 (2009).

20. Park, J. S. *et al.*, G' band Raman spectra of single, double and triple layer graphene. *Carbon* 47, 1303 (2009).

21. Malard, L. M., Mafra, D. L., Doorn, S. K. & Pimenta, M. A., Resonance Raman scattering in graphene: Probing phonons and electrons. *Solid State Commun.* 149, 1136–1139 (2009).

22. Mafra, D. L. *et al.*, A study of inner process double-resonance Raman scattering in bilayer graphene. *Carbon* 49, 1511–1515 (2011).

23. May, P. *et al.*, Signature of the two-dimensional phonon dispersion in graphene probed by double-resonant Raman scattering. *Phys. Rev. B* 87, 075402 (2013).

24. Pisana, S. *et al.*, Breakdown of the adiabatic Born-Oppenheimer approximation in graphene. *Nat. Mater.* 6, 198–201 (2007).

25. Yan, J., Zhang Y., Kim, P., & Pinczuk, A., Electric field effect tuning of electron-phonon coupling in graphene. Phys. Rev. Lett. 98, 166802 (2007).

**Acknowledgements**

This work was supported by the National Research Foundation (NRF) grants funded by the Korean government (MSIP) (Nos. 2011-0013461 and 2011-0017605) and by a grant (No. 2011-0031630) from the Center for Advanced Soft Electronics under the Global Frontier Research Program of MSIP.

**Author Contributions**

D.Y. and H.C. conceived the experiments. T.A.N. and J.-U.L. performed sample preparation and carried out experiments as well as analyses of the data. The results were discussed by all the authors, and the manuscript was written through contributions of all authors. All authors have given approval to the final version of the manuscript.

**Additional Information**

**Supplementary information** accompanies this paper at
http://www.nature.com/scientificreports

**Competing financial interests:** The authors declare no competing financial interests.

License:

How to cite this article:

**Figure Legends**

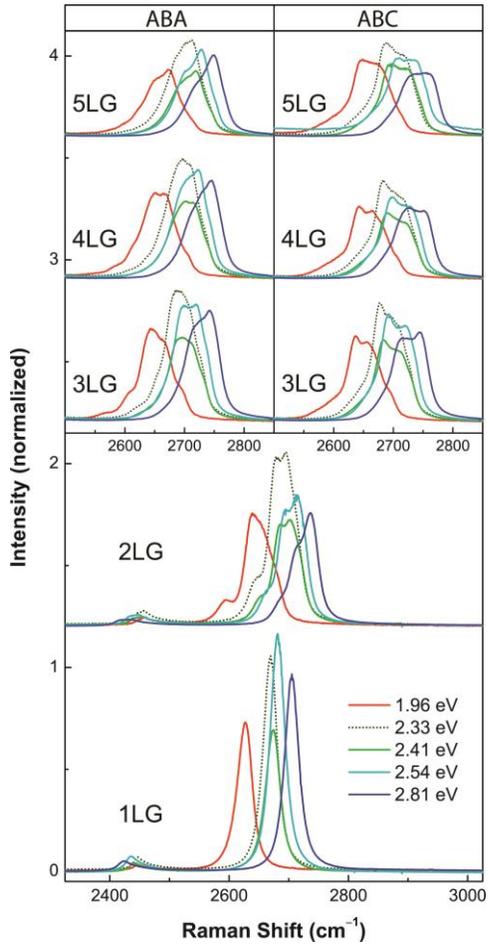

**Figure 1 | 2D band of FLG with 2 different stacking orders under 5 different excitation energies.** The blue, cyan, green, broken-green, and red lines represent the measured Raman spectra using 5 different excitation energies of 2.81, 2.54, 2.41, 2.33, and 1.96 eV, respectively. The 2D peak intensity is normalized to the G peak intensity for each spectrum. With increasing excitation energy, the 2D band blueshifts and changes its line shape. The line shapes of the 2D band are distinct between ABA- and ABC-stacking orders for all excitation energies.

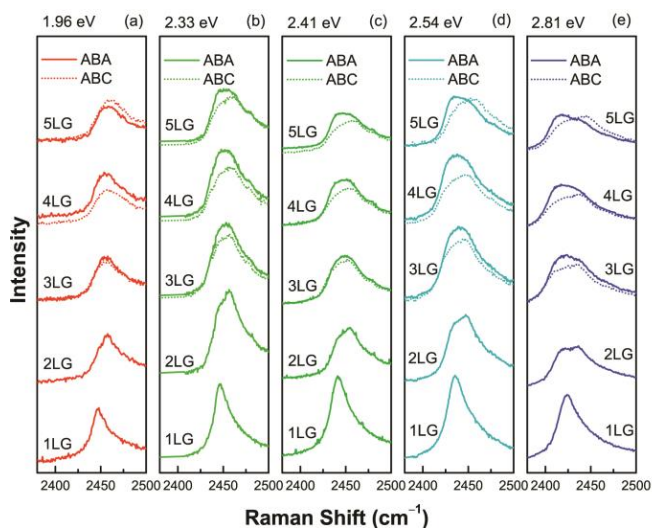

**Figure 2 | G\* band of FLG for excitation energies of (a) 1.96, (b) 2.33, (c) 2.41, (d) 2.54, and (e) 2.81 eV. The solid spectra are from monolayer and ABA-stacked graphene. The dotted spectra are for ABC-stacked FLG. The G\* band redshifts when excitation energy increases. We observed the difference in G\* band between ABA- and ABC-stacked FLG when 441.6 nm or 488 nm lasers were used.**

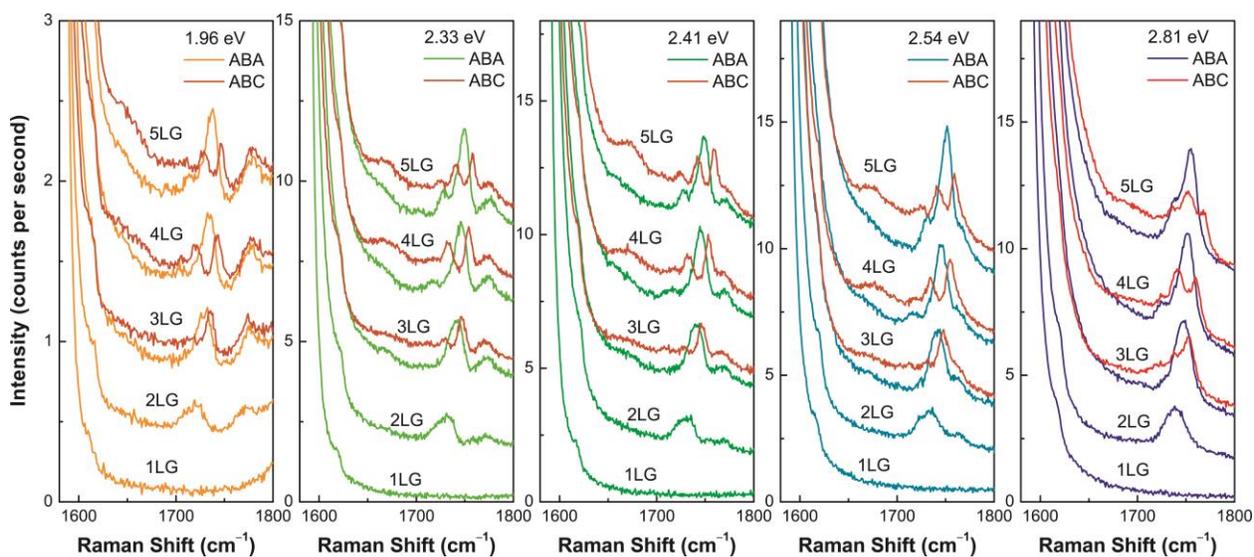

**Figure 3 | M band in FLG measured under 5 different excitation energies.**

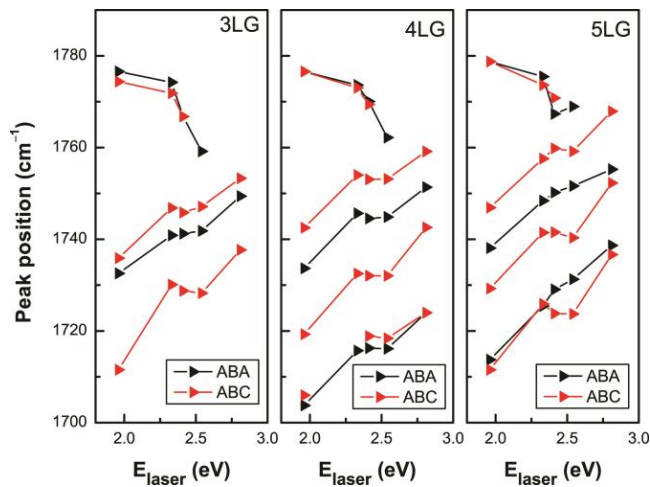

**Figure 4 | Dependence of the peak positions in the M band of ABA- and ABC-stacked FLG on the excitation energy.**

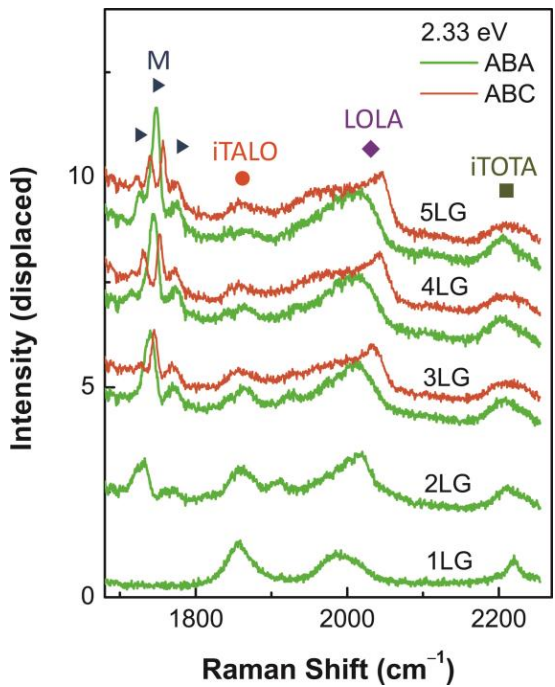

**Figure 5 | Combination modes in the range of 1700 to 2250 cm$^{-1}$ for the excitation energy of 2.33 eV. Green spectra are for single-layer and ABA-stacked FLG. Red spectra are for ABC-stacked FLG. ABA- and ABC-stacked FLGs show different M-band peaks and LOLA-mode peaks. Spectra taken with other excitation energies are shown in Supplementary Fig. S7.**

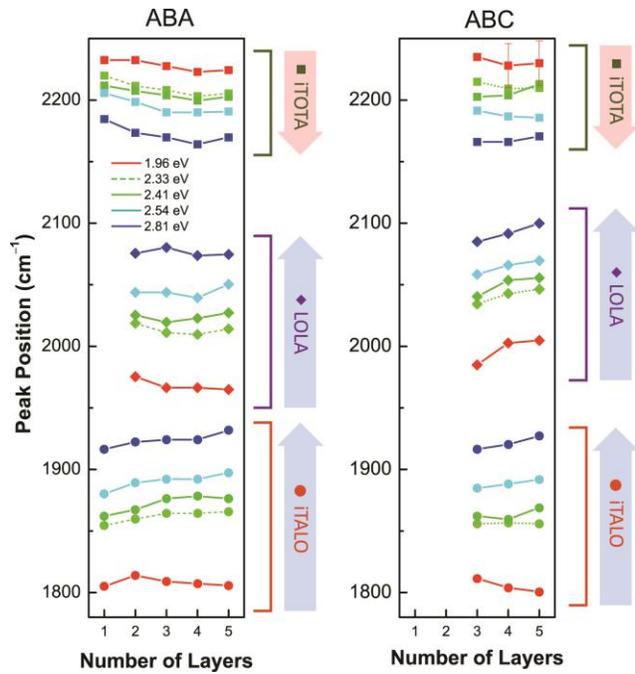

**Figure 6 | Dependence of the peak positions of the combination modes on the number of layers in ABA-stacked graphene in the range of 1780 to 2250 cm$^{-1}$.** The blue, cyan, green, dotted- green and red lines represent the measured peak positions of combination modes using 5 different excitation energies of 2.81, 2.54, 2.41, 2.33 and 1.96 eV, respectively. The arrows indicate the shifting direction of combination-mode peaks when the excitation energy increases.

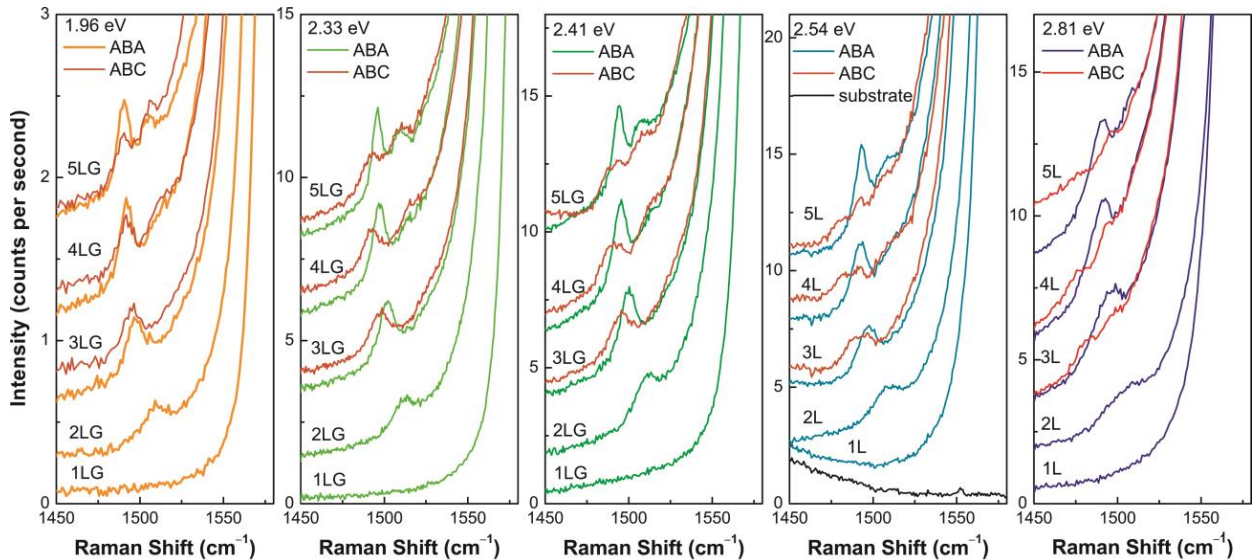

**Figure 7 | N band in FLG under excitation of 5 different lasers.** The broad background spectrum, which comes from the Si/SiO$_2$ substrate, is subtracted in the cases of 1.96, 2.33, 2.41, and 2.81 eV. In the cases of 2.54 eV, the background (black spectrum) was not removed so that the N band is better resolved.

# Excitation Energy Dependent Raman Signatures of ABA- and ABC-stacked Few-layer Graphene

The An Nguyen[1], Jae-Ung Lee[1], Duhee Yoon[1,†], and Hyeonsik Cheong[1,*]

[1]Department of Physics, Sogang University, Seoul 121-742, Korea.

[†]Current address: Cambridge Graphene Centre, Engineering Department, University of Cambridge, Cambridge, CB3 0FA, United Kingdoms.

1. **Schematics of vibration modes in 3LG**

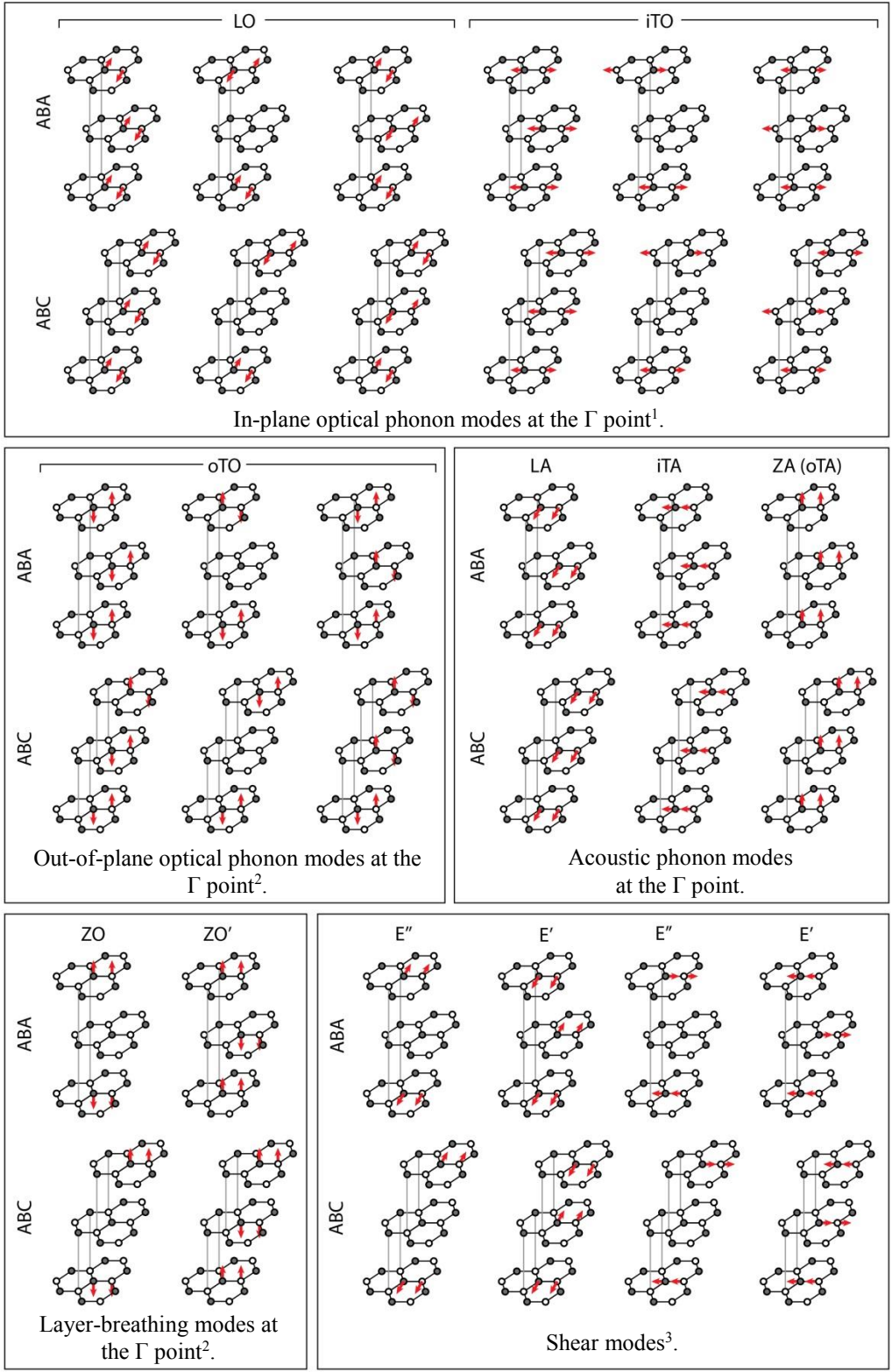

**Figure S1. Schematics of vibration modes in 3LG.**

## 2. Visualizing areas with different stacking orders by using Raman imaging technique

Since Bernal (ABA) and rhombohedral (ABC) stacked few-layer graphene (FLG) have distinct Raman 2D line shapes, we can exploit this feature to visualize areas of FLG samples which have different stacking orders[4]. Compared to ABA-stacked FLG, the Raman 2D band in ABC-stacked FLG has a more asymmetric shape and an enhanced peak and a shoulder. As a consequence, by fitting the 2D band into a single Lorentzian peak, one can expect the widths and peak positions of the fitted peaks to be different between these 2 stacking orders. A study by Lui *et al.* and our results agreed that the *single*-Lorentzian fitted 2D bands in ABA-stacked FLG samples have narrower peak 'widths' and blueshifted peak positions compared to samples which are ABC-stacked [Fig. S2(a)][4]. By imaging such 'width' and 'position', one can visualize domains with different stacking orders. Figures S2(b) and (c) show an optical image of a graphene sample and its image of 2D-peak 'width' as determined by the *single*-Lorentzian fitting. The optical image shows regions in uniform colors, indicating the same thicknesses. However, the Raman image shows domains with different stacking orders in one sample.

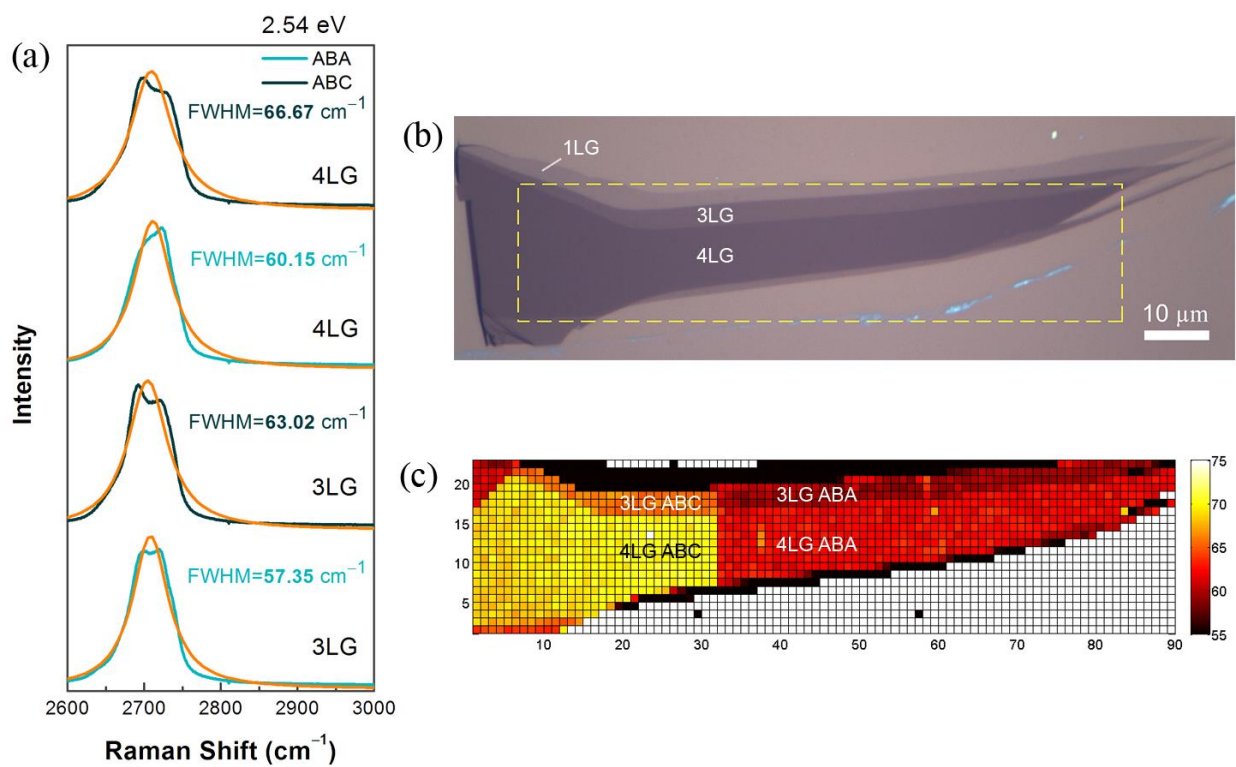

**Figure S2.** (a) Raman 2D bands of 3LG and 4LG with 2 different stacking orders (cyan and dark blue spectra) and their *single*-Lorentzian-fitted peaks (orange lines). (b) Optical image of a graphene sample. (c) Image of the 2D-peak 'width' of the same sample in (b). The yellow domain on the left-hand side is 4LG with the ABC-stacking order, and the red domain is 4LG with the ABA-stacking order. Near the top of this sample, there are narrow areas of ABA- (orange) and ABC-stacked 3LG (dark red).

### 3. Distinguishing stacked multilayer graphene and twisted/folded graphene.

During our investigation on ABA- and ABC-stacked graphene, we also encountered several incommensurate FLG samples. The misorientation or twist between two or more layers can modify the electronic structure[5-8]. The Raman spectra of such twisted FLG show some distinctive features, such as enhanced G and 2D band intensities or appearance of a small extra peak on the lower side of the G peak[9-12].

We compared the Raman 2D bands between folded FLG and stacked FLG. Figure S3(a) shows an optical image of a graphene sample in which an area of AB-stacked 2LG folds on domains of 2LG and 1LG. Figure S3(b) shows another folded graphene sample, in which a region of 3LG with the ABA stacking order folds onto itself. The Raman spectra of these folded graphene and stacked graphene [Figure S3(c), (d)] were measured under the same condition. We observed that under the same measurement conditions, the intensities of the 2D peaks in folded graphene are always several times larger than in FLG which has a stacking order. Additionally, the 2D line shapes of folded FLG do not resemble the line shapes of stacked FLG despite their same thicknesses. For example, Fig. S3(d) shows that the Raman 2D band of folded 3LG (3LG on 3LG) does not share similarity with the 2D band of ABA-stacked 6LG, in terms of intensity and line shapes.

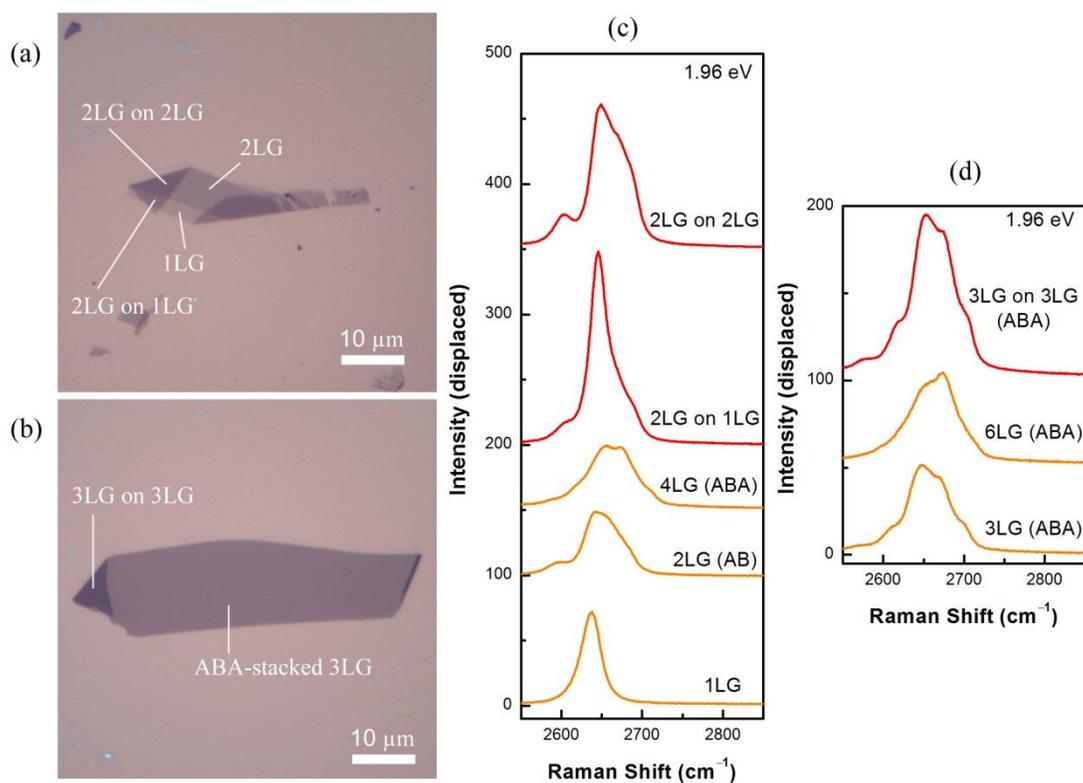

**Figure S3. (a, b)** Optical images of folded graphene samples on $SiO_2$/Si substrates. **(c, d) Raman 2D bands of ABA-stacked FLG (orange lines) and folded FLG (red lines) were measured in the same condition with the excitation energy of 1.96 eV. The intensities of the folded-graphene samples are much larger than those of ABA-stacked FLG.**

Beside the 2D band, we measured the Raman combination modes in the range from 1700 to 2250 cm$^{-1}$ of folded FLG. We observed no blueshift of the LOLA-mode peak (Figure S4). As explained in the main report, the blueshift of the LOLA-mode peak occurs in ABC-stacked FLG but not in ABA-stacked FLG. If the excitation energy is 2.41 eV, the Raman spectra of ABC-stacked FLG will show a LOLA-mode peak at 2057 cm$^{-1}$. There is no sharp peak at 2057 cm$^{-1}$ in the spectra of folded FLG as we have seen in the cases of ABC-stacked FLG. This means that the twisting or mismatching between graphene layers does not cause the blueshift of the LOLA-mode peak.

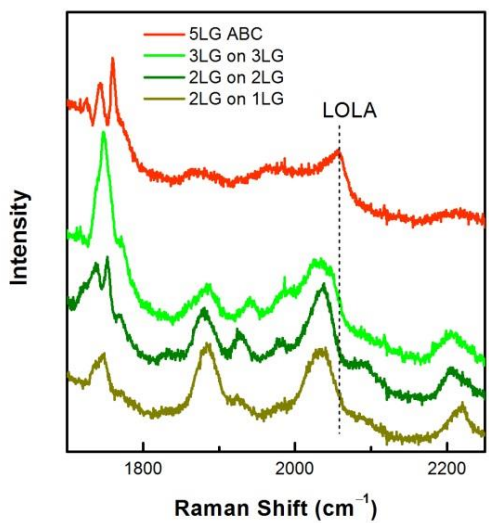

**Figure S4. Combination Raman modes in the range between 1700 and 2250 cm$^{-1}$ of folded graphene (green spectra) and ABC-stacked 5LG (red spectrum) for the excitation energy of 2.41 eV. The blueshifted LOLA-mode peak appears in ABC-stacked FLG only.**

## 4. Excitation Energy Dependence of the Raman 2D band of FLG

The line shapes of the Raman 2D bands in ABA- and ABC-stacked FLG for different excitation energies are summarized in Fig. S5. These are the same data as those presented in Fig. 1 of the main paper but arranged differently to allow easy comparison between different excitation energies.

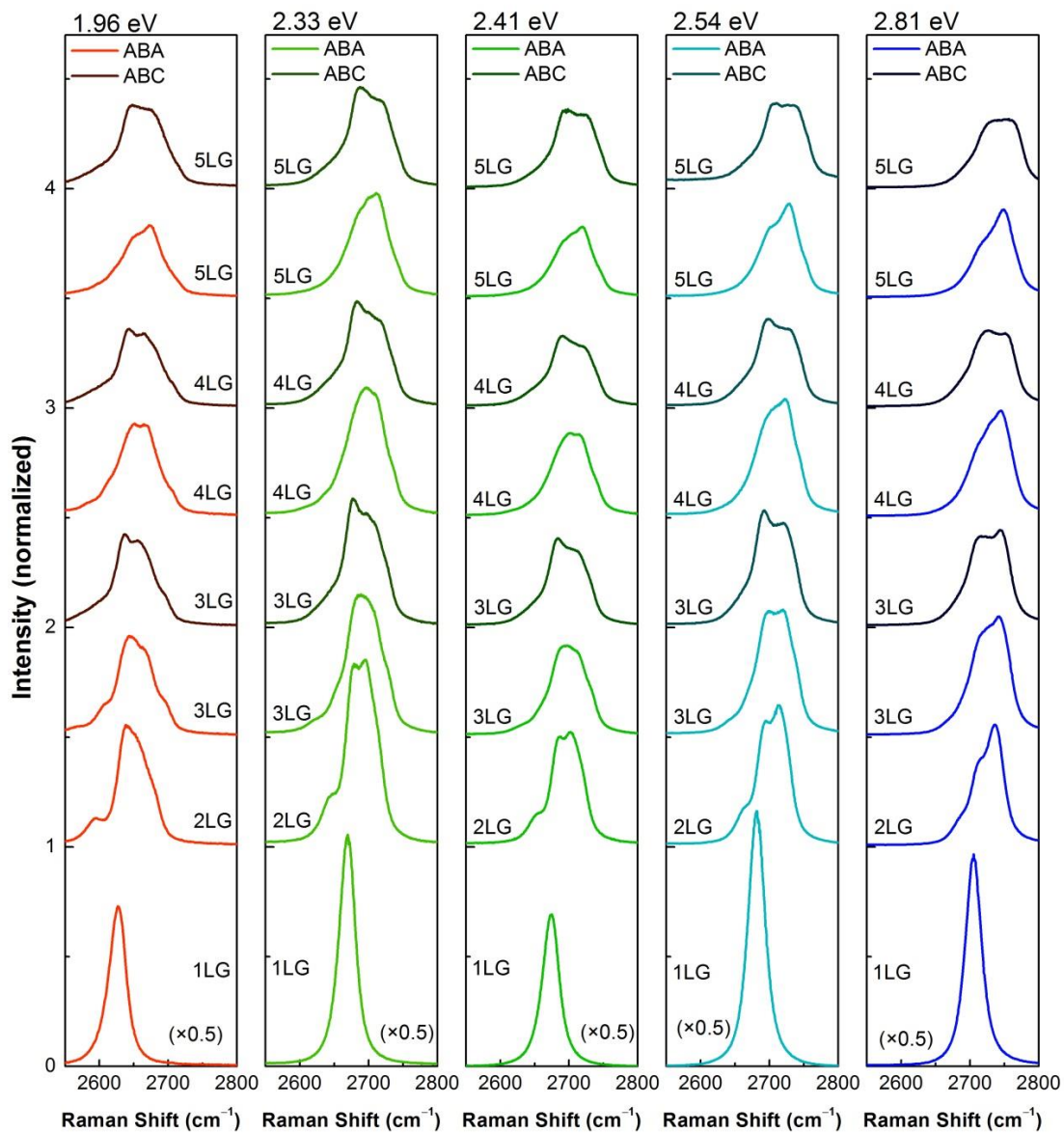

**Figure S5. Raman 2D bands in ABA- and ABC-stacked FLG. The 2D peak intensity is normalized to the G peak intensity for each spectrum.**

## 5. Raman spectra of combination modes from 1450 to 2250 cm$^{-1}$

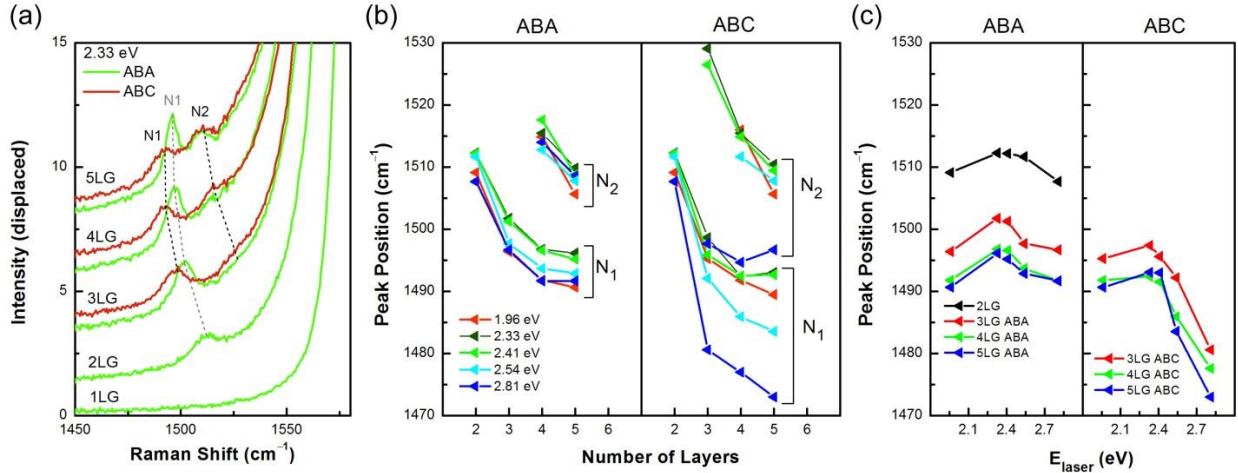

**Figure S6.** (a) Raman N-mode in FLG for an excitation energies of 2.33 eV. Green spectra are from monolayer and ABA-stacked FLG. Red spectra are for ABC-stacked FLG. Spectra from other excitation energies are shown in Fig. 7 of the main paper. (b) Dependence of the N-mode peak positions on the number of layers and the stacking order. (c) Excitation energy dependence of N$_1$ peak position in ABA- and ABC-stacked FLG.

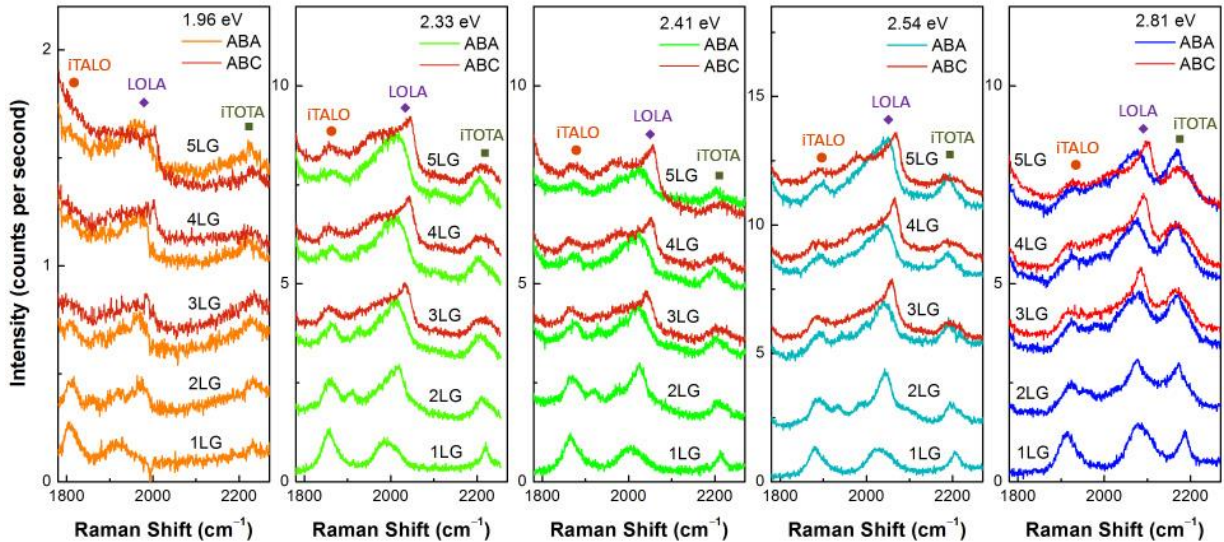

**Figure S7.** Combination Raman modes from 1780 to 2270 cm$^{-1}$ in FLG, measured with 5 different excitation energies. Most of the peaks are similar between ABA- and ABC-

**stacking orders, except the LOLA peak. The LOLA peak in ABC-stacked FLG is sharper and stays at higher frequency side compare to that in ABA-stacked FLG.**

**REFERENCES**

1. Yan, J.-A., Ruan, W. Y. & Chou, M. Y., Phonon dispersions and vibrational properties of monolayer, bilayer, and trilayer graphene: Density-functional perturbation theory. *Phys. Rev. B* **77**, 125401 (2008).
2. Cong, C. *et al.*, Raman Characterization of ABA- and ABC-Stacked Trilayer Graphene. *ACS Nano* **5** (11), 8760–8768 (2011).
3. Tan, P. H. *et al.*, The shear mode of multilayer graphene. *Nature Materials* **11**, 294-300 (2012).
4. Lui, C. H. *et al.*, Imaging Stacking Order in Few-Layer Graphene. *Nano Lett.* **11**, 164-169 (2011).
5. Novoselov, K. S. *et al.*, Two-dimensional gas of massless Dirac fermions in graphene. *Nature* **438** (10), 197-200 (2005).
6. Santos, J. M. B. L. d., Peres, N. M. R. & Neto, A. H. C., Graphene Bilayer with a Twist: Electronic Structure. *Phys. Rev. Lett.* **99**, 256802 (2007).
7. Laissardière, G. T. d., Mayou, D. & Magaud, L., Localization of Dirac Electrons in Rotated Graphene Bilayers. *Nano Lett.* **10**, 804-808 (2010).
8. Shallcross, S., Sharma, S., Kandelaki, E. & Pankratov, O. A., Electronic structure of turbostratic graphene. *Phys. Rev. B* **81**, 165105-165120 (2010).
9. Yoon, D., Cheong, H., Choi, J. S. & Park, B. H., Enhancement of the Raman scattering intensity in folded bilayer graphene. *J. Korean Phys. Soc.* **60** (8), 1278-1281 (2013).
10. Kim, K. *et al.*, Raman Spectroscopy Study of Rotated Double-Layer Graphene: Misorientation-Angle Dependence of Electronic Structure. *Phys. Rev. Lett.* **108** (24), 246103 (2012).
11. Podila, R., Rao, R., Tsuchikawa, R., Ishigami, M. & Rao, A. M., Raman Spectroscopy of Folded and Scrolled Graphene. *ACS Nano* **6**, 5784-5790 (2012).
12. Havener, R. W., Zhuang, H., Brown, L., Hennig, R. G. & Park, J., Angle-Resolved Raman Imaging of Interlayer Rotations and Interactions in Twisted Bilayer Graphene. *Nano Lett.* **12** (6), 3162-3167 (2012).